\documentclass[prl,preprint,showpacs]{revtex4}
\usepackage{epsfig}
\usepackage{graphicx}
\usepackage{amssymb}

\begin{document}

\title{Should we abandon cascade models to describe the spatial complexity of
fully developed turbulence velocity profiles ?}
\author{ J. Delour, J.F. Muzy and A. Arneodo} 
\address{Centre de Recherche Paul Pascal, Avenue Schweitzer, 33600 Pessac, France}
\date{\today}

\begin{abstract} 
We perform one- and two-points magnitude cumulant analysis
of one-dimensional longitudinal
velocity profiles stemming from three different experimental
set-ups and covering a broad range of Taylor scaled Reynolds numbers from
$R_\lambda =$ 89 to 2500. 
While the first-order cumulant behavior is found to strongly
depend on Reynolds number and experimental conditions, the second-order
cumulant and the magnitude connected correlation functions
are shown to display respectively universal scale and space-lag behavior.
Despite the fact that the Extended Self-Similarity 
(ESS) hypothesis is not consistent with these findings, 
when extrapolating our results to the limit of
infinite Reynolds number, one confirms the validity of the log-normal
multifractal description of the intermittency phenomenon 
with a well defined
intermittency parameter $C_2 = 0.025 \pm 0.003$. 
But the convergence to zero of the magnitude connected correlation 
functions casts doubt on the asymptotic existence of an underlying
multiplicative cascading spatial structure.

\end{abstract}


\maketitle

Since Richardson's original work \cite{bRic22}, a common ``mental image'' of
fully developed turbulence is a dynamical cascading process in which large
eddies split up into smaller ones which themselves blow up into even smaller
ones and so forth \cite{bFri95}. According to this picture, energy propagates
from the integral scale, where eddies
are generated, down to the dissipative scale, where they vanish by viscous
dissipation, through a multiplicative process, each eddy inheriting a fraction
of its parent's energy. Since this early intuitive description, the notion
of {\em cascade} has remained the creed of many models proposed in the
literature \cite{bFri95} to mimic the statistical properties of turbulent
signals. Among the available experimental data, a lot of effort has been
devoted to the study of the longitudinal velocity component recorded in
directional flows such as jets or wind tunnels
\cite{bFri95,bMon75,Ans84,Arn96}. In these configurations, the Taylor
hypothesis that considers
the spatial structure of the flow as globally advected upon the probe,
enables us to interpret temporal time series as spatial profiles.
In 1941, Kolmogorov \cite{Kol41} resumed Richardson's picture in his
statistical analysis of the spatial fluctuations of velocity profile in the
sense that he linked the one-point statistics of the velocity increments
$\delta v_l = v(x+l)-v(x)$ over different distances $l$, by some dimensional
analysis which predicted the remarkable scaling behavior of the moments
of $\delta v_l$ : $M_{q,l} = <\delta v_l^q> \sim l^{\zeta_q}$ where $\zeta_q = 
q/3$. Actually, $\zeta_q$ turned out to be a non-linear function of $q$ in most
experiments \cite{bFri95,bMon75,Ans84,Arn96} and many studies inspired from
Kolmogorov and Obukhov second theory \cite{Kol62} tried to explain and to
predict the analytical shape of this non-linearity. The controversial situation
at the origin of some disagreement between models ({\em e.g.} the
log-normal \cite{Kol62,Cas90,Arn98a} as opposed to log-Poisson models
\cite{Dub94}) results from the experimental observation that the moments
$M_{q,l}$ do not really scale perfectly. Indeed, there is a persistent
curvature when one plots $\ln(M_{q,l})$ {\em vs} $\ln(l)$, which means that,
rigorously speaking, there is no scale invariance. In order to give a sense to
the exponents $\zeta_q$, Benzi {\em et al.} \cite{Ben95} defined the ``Extended
Self-Similarity'' (ESS) hypothesis by proposing the following behavior for the
moments of the velocity increments :
\begin{equation}
\label{ESS}
M_{q,l} \sim f(l)^{\zeta_q},
\end{equation}
where $f(l)$ would be some $q$-independent function of $l$.
Along this line, $\ln(M_{q_1,l})$ {\em vs} $\ln(M_{q_2,l})$ curves look
definitely more linear than using standard log-scale representations and, by
assuming that $\zeta_3 = 1$ \cite{bFri95}, some experimental consensus has
apparently been reached on the nonlinearity of the $\zeta_q$ spectrum
\cite{Arn96}. In a recent theoretical work \cite{aArad99}, Arad {\em et al.}
suggest that at low Reynolds number, structure functions scaling properties are
``poluted'' by anisotropic effects that can be mastered using the irreductible
representations of the rotation group. Unfortunately, this analysis is not
tractable with single point data. Hopefully, as shown below, some statistical
quantities turn out to display universal behavior that are likely to be
insensitive to anisotropic effects.

In the early nineties, Castaing {\em et al.} \cite{Cas90} recasted the cascade
picture and the ESS hypothesis in a probabilistic description that accounts for
the continuous deformation of the probability density functions (pdf) of
$\delta v_l$ with $l$, by the mean of a propagator $G_{l l'}$ ($l'>l$): 
\begin{equation}
\label{Castaing1}
P_l(\delta v) = \int_{-\infty}^{\infty} G_{l l'}(u)e^{-u}P_{l'}(e^{-u} \delta v
) du.
\end{equation}
This equation can be related to a cascade process in which the variable
$\delta v_l$ is continuously decomposed 
as $\delta v_l = \prod_{i=1}^{n} W_{l_{i+1},l_i} \delta v_{l'}$,
where $\ln(|W_{l_{i+1},l_i}|)$ are independent random variables of law
$G_{l_{i+1}l_i}$.
Within this framework, the $\zeta_q$ and the $f(l)$ ESS functions 
can be related to the shape of $G_{ll'}$ thanks to the cumulant
generating function of $G$ \cite{Cas95,Arn98a}, 
\begin{equation}
\label{cumexp}
\ln \hat G_{ll'}(-iq) =  \ln \left( M_{q,l}/M_{q,l'} \right) = \zeta_q 
	\left(f(l)-f(l')\right), 
\end{equation}
the classical scale invariant case corresponding to $f(l) = \ln(l)$.
In this paradigm, the successive terms of the polynomial
development of $\ln \left( M_{q,l}/M_{q,l'} \right)$ as a function of $q$,
involve the cumulants $C_q(l)= - C_q f(l)$ of the logarithm of
$|\delta v_l|$ (``magnitude'') from which one can express the $\zeta_q$
spectrum as $\zeta_q = - \sum^{+\infty}_{k=1} C_k \frac{q^k}{k!}$.

The question whether turbulent velocity signals do (or do not)
present a space-scale cascade structure thus amounts to 
checking that the successive velocity magnitude cumulants
possess the same scale behavior. This 
perspective can be extended from structure functions
to unfused correlations functions \cite{Lvo96} (e.g., 
$<\delta v_l^q(x) \delta v_{l'}^p(x+\Delta x)>$), by studying the behavior 
of the magnitude connected correlation functions (MCCF) :
$C_{ll'}(\Delta x) = \linebreak
<\ln(|\delta v_l (x)|) \ln(|\delta v_{l'} (x+\Delta x)|)>_c$.  
These correlation functions have been introduced in 
Ref. \cite{Arn98c} and play the same role for the
multivariate multi-point velocity law as the previous magnitude cumulants
for the one-point law. For multiplicative cascade processes, they are expected
to display a logarithmic dependence on the spatial distance $\Delta x$
\cite{Arn98c}. Our goal in this Letter is to push further the analysis carried
out in Refs \cite{Arn98a,Arn98c} by studying both the cumulants and the
connected 
correlation functions associated to $\ln(|\delta v_l|)$ at different scales
$l$, with the aim to establish a clear diagnostic about (i) the scaling
properties of experimental turbulent velocity records and (ii) the validity of
multiplicative cascade models in the limit of infinite Reynolds number.
We will work with signals gracefully supplied by three different
experimental groups. The signals labelled with Taylor scaled Reynolds number
$R_\lambda =$ 89, 208, 463, 703 and 929 stem from Castaing's group and were
recorded in a gaseous helium jet at very low temperature in CRTB (Grenoble)
\cite{thCha}; the signal labelled $R_\lambda =$ 570 stems from Baudet's group
and was recorded in an air jet in LEGI (Grenoble);
the signal labelled $R_\lambda =$ 2500 was recorded at the ONERA wind tunnel in
Modane by Gagne and collaborators \cite{thMal}.

Let us start investigating one-point statistics via the computation of the
three first cumulants $C_q(l)$ of $\ln(|\delta v_l|)$. In Fig.~1a, $C_1(l)$ is
plotted {\em vs} $\ln(l)$ for all signals. They are all characterized by an
integral scale $L$ corresponding to a decorrelation of the increments
$|\delta v_l|$ for $l>L$. Below this saturated regime, 
there is no well defined ``inertial range'' since one observes a continuous
cross-over across scales towards the smooth dissipative regime down the
Kolmogorov scale $\eta$.
Let us note that the integral scale does not
depend on the Reynolds number but only on geometrical considerations. As an
illustration, one gets the same value for $L$ for all the signals stemming from
Castaing's group which were recorded in the same apparatus, $R_\lambda$ being
controlled either by the mean speed of the flow or by the value of
viscosity that is changed by tuning temperature. Furthermore, the numerical
values obtained for $L$ : $1.82$ cm for Castaing's group, 0.91 m
for Baudet's group and 10.2 m for Gagne's group, correspond to the
characteristic sizes of the experimental set-ups. $L$ is thus a geometric
parameter and when one increases $R_{\lambda}$,
the inertial range increases via the decrease of $\eta$. 
Let us notice that as proposed in Ref. \cite{thCha},
the shape of the cross-over from the saturated regime to the inertial one
is rather well reproduced by a simple Langevin equation (inset of Fig.~1d) :
$\delta v(x) = - \gamma v(x) + \sigma W(x)$, where
$\gamma = 1/L$ and $W(x)$ is an appropriate fractional Brownian noise ($B_{H}
= 1/3$). However the precise functional dependence of $C_1(l)$ 
is likely to depend on the experimental parameters and
to be strongly influenced by anistropic effects \cite{aArad99}.

The situation is very different for the second magnitude cumulant $C_2(l)$ when
plotted {\em vs} $\ln(l)$ as shown in Fig.~1b. We first notice that the
decorrelation scale where $C_2(l)$
saturates, is even better defined with $C_2(l)$ than with $C_1(l)$, since a
linear behavior is observed up to this scale without any cross-over. The most
important feature is that, whatever the set-up and whatever $R_{\lambda}$, all
the data fall on {\em linear} curves which all have the {\em same} slope
$- C_2 = - 0.025 \pm 0.003$. As a matter of fact, if we superimpose
all the curves by setting all $L$ to 1, as shown in Fig.\,1e, we cannot
distinguish anymore one curve from the others. This consideration is
particularly relevant in the case of the signals of Castaing's group where
$R_\lambda$ is varied without changing $L$.

The results concerning the third magnitude cumulant $C_3(l)$ are reported in
Figs 1c and
1f. The slope of the curves obtained when plotting $C_3(l)$ {\em vs} $\ln(l)$
systematically decreases when increasing $R_\lambda$. For small $R_\lambda$,
$C_3$ is significantly different from zero which means that the log-normal
paradigm is not valid. In fact, from the curvature of the corresponding
experimental $C_1(l)$ curves, we believe that
this is rather a confirmation of the absence of a well defined inertial range
for these low values of $R_\lambda$. For the largest values of $R_\lambda$
($\geq 800$), $C_3$ becomes small enough (up to finite sample effects as
previously reported in Ref. \cite{Arn98a}) to be neglected \cite{note}. 
At high Reynolds numbers, one can thus suppose that $C_3=0$ (and thus 
so the higher order cumulants) that implies a normal shape for the
propagator $G_{ll'}$ in
agreement with log-normal models \cite{Kol62,Cas90,Arn98a}. To summarize, the
scale behavior of structure functions is well described by the first two terms
in the cumulant expansion of $G_{ll'}$ and the 
differences observed in the behavior of $C_1(l)$ (Figs 1a and 1d) and $C_2(l)$
(Figs 1b and 1e) bring the experimental demonstration of the inconsistency of
the ESS hypothesis which rigorously requires an identical scale behavior for
the two cumulants (Eq. (\ref{cumexp})). The full $\zeta_q$ spectrum cannot thus
be defined in the range of Reynolds numbers we have investigated but
the scaling exponent $C_2 = 0.025 \pm 0.003$ of $C_2(l)$ is likely to be a
universal (model independent) characteristic of intermittency.

Let us now focus on the MCCF : $C_{ll'}(\Delta x) =
<\ln(|\delta v_l (x)|) \ln(|\delta v_{l'} (x+\Delta x)|)>_c$. As pointed out
in Ref.\cite{Arn98c}, we have checked that the MCCF computed at different
scales $l$ and $l'$ all collapse on a single curve provided $\Delta x >
\max(l,l')$. Let us recall that for a multiplicative cascade, we expect the
MCCF to behave as $- C_2 \ln(\Delta x / L)$. In Fig.\,2 are shown the MCCF
computed at a single small scale
($l=l'$) for each experimental signal. Three main observations must be raised
from these curves. First, the MCCF do not behave linearly but as the square of
the logarithm of the spatial distance $\Delta x$ : $C_{ll} (\Delta x) =
\alpha(R_{\lambda}) \ln^2(\Delta x / L)$. This quadratic behavior is universal
as far as experimental set-ups are concerned and has never been observed
before. Second, the respective integral scales at which the magnitudes are
decorrelated are nearly the same as the ones observed on the corresponding
cumulants $C_q(l)$ ($q =$ 1,2), which is not a trivial result.
Finally, as shown in Fig.~3, the
prefactor $\alpha(R_{\lambda})$ has a systematic decreasing behavior as a
function of $R_\lambda$. As far as the analytic shape of this decrease
is concerned, there is no certainty. As illustrated in Fig.~3a, a power-law
behavior~: $\alpha \sim R_{\lambda}^{-a}$ with $a \simeq 0.20 \pm 0.03$
provides a
reasonable good fit of the data. But as shown in Fig.~3b, one can also fit
the data by $\alpha \sim K/\ln(R_{\lambda})$ with $K = 0.027 \pm 0.006$, which
is a particular Reynolds number dependance that allows us to define a
characteristic
scale of the flow under study. Indeed, if one considers that the decorrelation
scale is the same for the magnitude cumulants $C_q(l)$ and the MCCF, from the
behavior of the variance as $-C_2 \ln(l / L)$ and of the MCCF as $\alpha
\ln^2(\Delta x / L)$, one can define a characteristic scale $l_c$ where the two
curves meet (when identifying scale and space-lag according to the
multiplicative cascade picture \cite{Arn98c}) : $\ln(l_c/L) =
-C_2/\alpha$. With the numerical values extracted from Figs 1b, 2 and 3b,
one gets $(l_c/L) \sim R_\lambda^{-b}$ with $b = 0.91 \pm 0.3$. This new scale
$l_c$ might well be the
Taylor scale $\lambda$ for which $b = 1$. In that respect, our approach
could be a way to measure objectively $\lambda$ and this even when the
dissipative scale $\eta$ is not resolved. This universal quadratic behavior of
the MCCF {\em vs} $\ln(\Delta x)$ and the systematic decrease of
$\alpha(R_{\lambda})$ when increasing $R_{\lambda}$, are new observations which
enables us to make some conjecture about the asymptotic limit of infinite
Reynolds number. As pointed out in Refs \cite{Arn98a,Arn98c}, correlations in
the magnitude are the signature of a multiplicative structure. If we
extrapolate the observed behavior to infinite $R_{\lambda}$, we predict the
convergence to zero of the MCCF on any finite range of scales. That is to say,
in the limit of infinite $R_{\lambda}$, the intermittent spatial fluctuations
of the longitudinal velocity component are not likely to display a
multiplicative hierarchical structure.

To summarize, we have advocated in this paper the
study of scaling properties of longitudinal velocity increments
by means of one- and two-points magnitude cumulant analysis.
The results of our measurements for seven flows, at
seven different $R_\lambda$ and stemming from different experimental set-ups,
are two-fold. Concerning the one-point (fused) statistics,
we mainly observe that the log-normal approximation 
is pertinent at sufficiently high Reynolds number
and that the first two cumulants have not the same scale
behavior as assumed by ESS hypothesis.
While $C_1(l)$ displays some Reynolds and set-up dependent departure from scale
invariance, $C_2(l)$ exhibits scale invariance behavior with a universal
intermittency coefficient $C_2 = 0.025 \pm 0.003$.
Concerning the two-points (unfused) statistics, we observe that
the behavior of the MCCF {\em vs} $\ln(\Delta x)$ is
quadratic whatever the considered experiment but with a prefactor
$\alpha(R_\lambda)$ which decreases when increasing $R_\lambda$.
These new observations lead us
to conjecture that in the limit of infinite Reynolds number, the space-scale
structure underlying the fluctuations of the longitudinal velocity is not of a
multiplicative type but more likely in the spirit of the log-normal
multifractal description pioneered by Kolmogorov and Obukhov in 1962
\cite{Kol62}, i.e. with an intermittency coefficient $C_2 = 0.025
\pm 0.003$ but without any correlations across scales. This is not at all
shocking from a physical point of view since the ``mental image'' of a
dynamical cascading process \cite{bFri95} does not {\em a priori} imply that
there should be some multiplicative (cascading) spatial organization.
Furthermore, one must realize that our conclusions come out from the study of
1D cuts of the velocity field only. In a forthcoming publication,
we plan to carry out a similar analysis of a 3D velocity field
issued from direct numerical simulations, with the specific goal of testing the
validity of this non multiplicative log-normal multifractal picture to account
for the intermittent nature of fully developed turbulent 3D velocity fields.

We are very grateful to B.Castaing's, C.Baudet's and Y.Gagne's groups for the
permission to use their experimental signals.

\begin{figure}[t]
 \begin{center}
    \epsfig{file=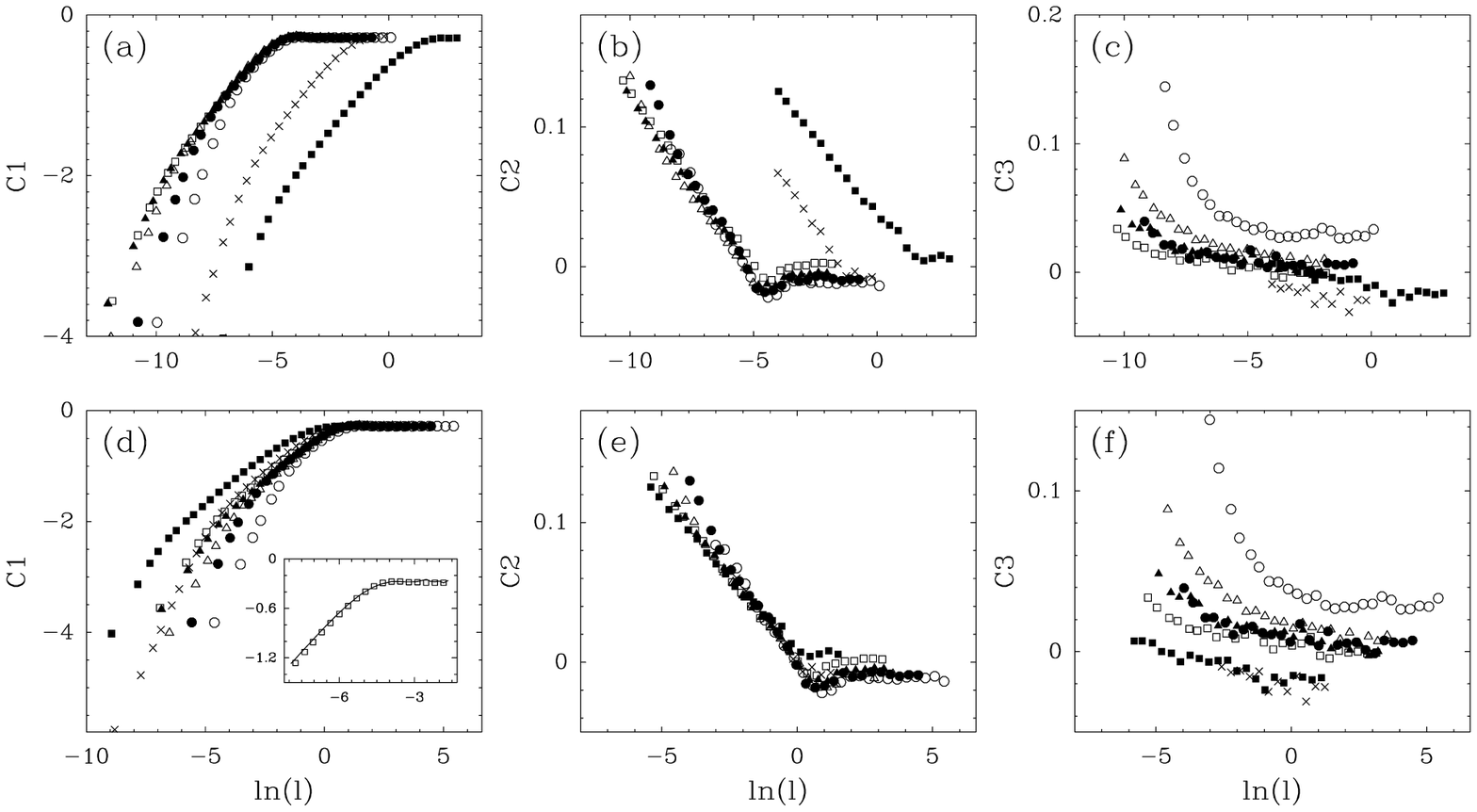,width=16.5cm}
    \caption{Magnitude cumulants $C_q(l)$ of
the seven studied signals for $R_\lambda =$ 89 ($\circ$), 208 ($\bullet$), 463
($\vartriangle$), 570 ($\times$), 703 ($\blacktriangle$), 929 ($\square$) and
2500 ($\blacksquare$). (a) $C_1(l)$ {\em vs} $\ln(l)$ : some continuous
cross-over is observed from a decorrelated regime at large scales down to a
smooth one at small scales. (b) $C_2(l)$ {\em vs}
$\ln(l)$ : a scaling behavior is obtained with the same slope $-C_2 =
-0.025 \pm 0.003$ for all signals. (c) $C_3(l)$ {\em vs} $\ln(l)$ : in the
inertial range and for the largest values of $R_\lambda$, the slope is
compatible to zero up to finite sample effects. (d), (e) and (f) : same curves
as in (a), (b) and (c), when all integral scales are set to 1. In the inset of
(d), the solid line corresponds to the prediction of the Langevin model
defined in the text after adjusting the integral scale $L$ and the parameter
$\sigma$ to the measured ones.}
 \end{center}
\end{figure}

\begin{figure}[t]
  \begin{center}
    \epsfig{file=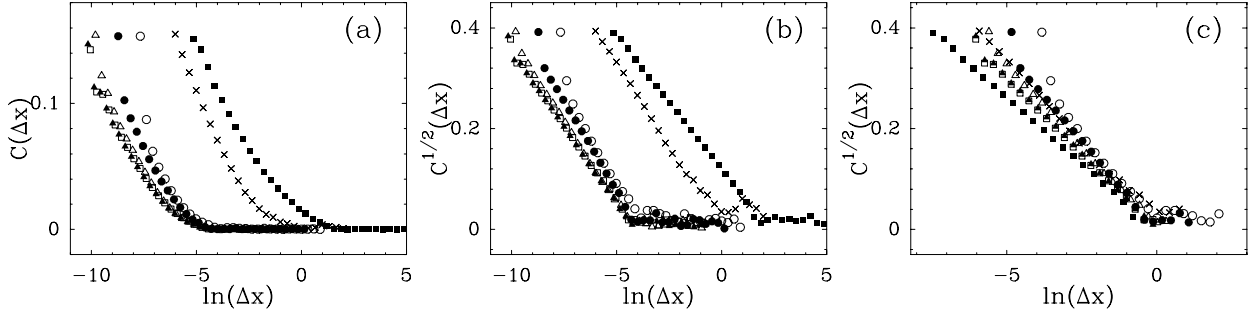,width=16.5cm}
    \caption{(a) Magnitude connected correlation functions $C_{ll}(\Delta x) =
<\ln(|\delta v_l (x)|)$ $\ln(|\delta v_l (x+\Delta x)|) >_c$ {\em vs}
$\ln(\Delta x)$. (b)
Square root of $C_{ll}(\Delta x)$ {\em vs} $\ln(\Delta x)$; $C_{ll}(\Delta x)$
behaves as $\alpha \ln^2(\Delta x)$ whatever the experimental set-up with a
systematic decrease of $\alpha$ when $R_\lambda$ is increased. (c) Same curves
when all integral scales $L$ are set to 1. The considered scale is $l \simeq$
50 $\mu$m for Castaing's signals, 0.15 mm for Baudet's signal and 3.2 mm for
Gagne's signal.
The symbols correspond respectively to the seven  same signals as in Fig.~1.} 
  \end{center}
\end{figure}

\begin{figure}[t]
  \begin{center}
    \epsfig{file=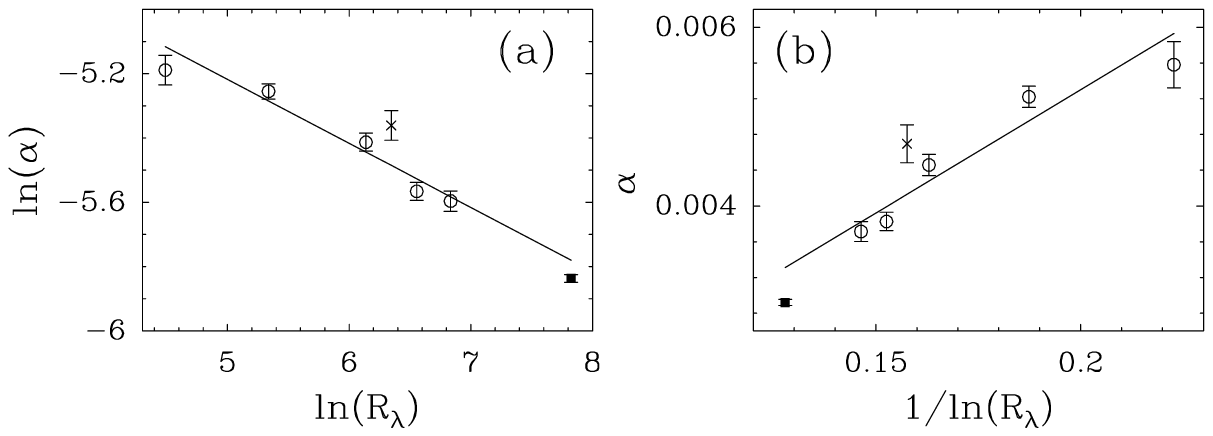,width=12cm}
    \caption{Prefactor $\alpha(R_\lambda)$ of the quadratic behavior of $C_{ll}
(\Delta x)$ {\em vs} $\ln(\Delta x)$ (see Fig.~2). (a) $\ln(\alpha)$ {\em vs}
$\ln(R_\lambda)$; the data are compatible with a power-law behavior
$\alpha \sim R_\lambda^{-a}$ with $a = 0.20 \pm 0.03$ (solid line).
(b) $\alpha$ as a function of $1/\ln(R_\lambda)$; the data are compatible with
a behavior $\alpha \sim K/\ln(R_\lambda)$ with $K = 0.027 \pm 0.006$ (solid
line). The symbols have the following meaning~: ($\circ$) Castaing's group,
($\times$) Baudet's group and ($\blacksquare$) Gagne's group.} 
  \end{center}
\end{figure}


\begin{thebibliography}{18}
\bibitem{note} Note that statistical convergence is faster in the estimation of $C_3$ when using the propagator as in Ref. \cite{Arn98a} than when computing directly $C_3(l)$ as reported in Fig. 1c.
\end{thebibliography}
\end{document}